\documentstyle[epsf,twocolumn,floats,prl,aps]{revtex}
\newcommand{\figwidth}{3.375in}
\begin{document}
\draft

\twocolumn[\hsize\textwidth\columnwidth\hsize\csname @twocolumnfalse\endcsname

\title{Monolayer Solid $^4$He Clusters on Graphite}
\author{M. E. Pierce and E. Manousakis}
\address{
Department of Physics and Center for Materials Research and Technology,
Florida State University, Tallahassee, FL 32306-4350
}
\date{\today}
\maketitle
\begin{abstract}
\noindent
In order to resolve the controversy about the low density region of the 
phase diagram of the $^4$He monolayer on graphite, we have undertaken a path 
integral Monte Carlo study of the system.
We provide direct evidence that the low density monolayer possesses 
solid clusters and a low density
vapor as opposed to the most recent proposal that the system is in a 
superfluid phase.  We further establish that the rounded heat capacity 
peaks observed at low densities are caused by melting of such solid 
clusters and are not associated with the suggested superfluid transition.
\end{abstract}
] 

Monolayer helium adsorbed on graphite has long proven to be a 
fascinating system for studying the growth and behavior of a quantum
film above an atomically ordered, uniform substrate and has
been used to investigate a number of nearly two-dimensional (2D) 
phenomena \cite{dash75,schick80,bruch97}.  
A very prominent feature of this layer is
a commensurate solid phase in which one-third of the available 
substrate adsorption sites are occupied
\cite{bretz77,greywall93,nielsen80,abraham87}.  
For densities above the commensurate, the system is known to pass
through a region of domain wall phases before forming an
incommensurate triangular solid phase.  On the other hand, the nature
of the phase diagram below the commensurate density at low 
temperatures is less well
established, with two competing pictures.  One possibility is that
the phase is a solid with vacancies \cite{ecke85}.  At low temperatures, the 
vacancies coalesce, producing coexistence between a solid cluster
and a vapor.  More recently, it has been suggested that the solid melts
if the density is decreased, and
there is a low temperature liquid phase \cite{greywall91}.
The first layer would then be a candidate in the search for a monolayer
superfluid \cite{greywall91,mono}.

The solid cluster picture has been 
discussed by Ecke {\em et al} \cite{ecke85}.
They note that since the commensurate solid
phase is in the same universality class as the three-state 
Potts model \cite{bretz73,berker78,schick77}, then at 
lower densities the film should consist of a commensurate solid
with vacancies.  If the temperature is raised, the solid 
melts continuously.  Lowering the temperature causes the vacancies to
coalesce (phase separate), a first order transition.  The difference 
between the temperatures of these two transitions becomes smaller
as the density is lowered until they meet at a tricritical point.  
This point was determined \cite{ecke85} to be at about
0.039 atom/$\AA^2$ and 1.3 K.  Thus, in this picture,
the monolayer consists of solid clusters surrounded by a low density vapor
at low temperatures and densities.

More recent experiments have questioned this conclusion.  Greywall
and Busch \cite{greywall91} point out that the heat capacity is not 
linear in density for the entire region below the commensurate density, 
as it must be for solid-vapor coexistence \cite{dash75}.  They instead
propose that the system has a self-bound liquid phase at about 0.04
atom/$\AA^2$.  This conclusion is supported by 2D variational 
calculations for helium \cite{bruch93} 
that take substrate corrugations into account.
The possibility of a first layer liquid is intriguing since at low
temperatures it would be a superfluid above a bare substrate, with
no underlying ``dead'' layer of helium.  However, direct 
measurements \cite{reppy96} on the first layer detect 
no superfluidity.  This negative result has been attributed to poor 
substrate connectivity.

In order to resolve this controversy, we have undertaken a path integral Monte
Carlo (PIMC) study of the low density first layer.  This is the first attempt
to directly investigate this region by an exact, first principles method that
treats the full quantum many-body problem.  From our calculations,
we provide the first direct evidence that the low density monolayer 
consists of solid clusters.  No liquid phase occurs, and so there is
no possibility for first layer superfluidity.  We further establish that
the rounded heat capacity peaks observed at low densities are caused by
the melting of solid clusters and are not associated with 
a superfluid film, as has been suggested \cite{greywall91}.  

Our PIMC calculations use realistic helium-helium \cite{aziz92}
 and helium-substrate
interactions.  In order to include the effects of substrate corrugations, we 
use the full, anisotropic helium-graphite potential of Ref. \cite{cole80}.
For a general discussion of the PIMC method and its application to 
films, see Refs. \cite{ceprev,pierce}.  
In test runs, we determined
that an inverse temperature slice of $\tau = 1/200 K^{-1}$ was required 
to reach the desired  
accuracy using the semiclassical approximation for the required
high temperature density matrix at 200 K.  The semiclassical 
approximation allows substrate
corrugations to be easily implemented.
From the same test runs, we determined that
an $l=3$  multilevel bisection was required.
All of our calculations are performed
in a simulation cell with periodic boundary conditions and 
dimensions $25.560 \AA \times 22.136 \AA$
that exactly accommodates the commensurate solid
The number of particles ranged from 20 to 40, with 36 corresponding
to the commensurate density, $\rho_c=0.0636$ atom/$\AA^2$.  We also allowed 
for the possibility of particle permutations at low densities, but
did not observe any.

It is essential
that corrugations be included in the calculations because the
commensurate solid phase is entirely
the result of the substrate corrugations.  A recent simulation
for the helium monolayer\cite{whitlock98} using the laterally 
averaged graphite potential\cite{cole80} finds that the equilibrium 
phase is a liquid.  
Full solidification does not occur until the coverage is well
above $\rho_c$.
At $\rho_c$, the film on the 
featureless substrate is a compressed uniform liquid that is
near the beginning of solid-liquid coexistence.

Before presenting evidence that the first layer has solid clusters, 
we first wish to demonstrate that our simulation method can
reproduce the commensurate solid phase, and that this phase
exhibits melting-like behavior in agreement with experiment.  We 
then investigate the 
low temperature phase diagram using the Maxwell construction.  

\begin{figure}[htp]
\epsfxsize=\figwidth\centerline{\epsffile{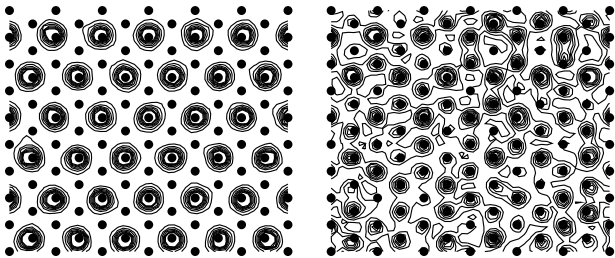}}
\caption{Distribution plots at the commensurate density, 0.0636 atom/$\AA^2$,
for  T=2.99 K (left) and T=4.0 K (right).  Filled circles indicate
graphite adsorption sites.
}
\label{fig:dist636}
\end{figure}

\begin{figure}[htp]
\epsfxsize=\figwidth\centerline{\epsffile{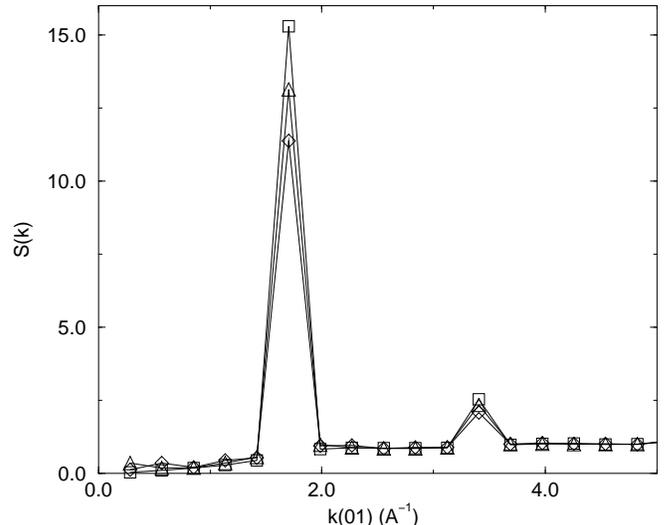}}
\caption{Static structure factor S(k) for 0.0530 (diamonds), 
0.0566 (triangles)
and 0.0636 atom/$\AA^2$ (squares) at 2.0 K.  The particle numbers are
are 30, 32, and 36, respectively.}
\label{fig:ssffull}
\end{figure}

Figure \ref{fig:dist636} illustrates the $\sqrt 3 \times \sqrt 3$
commensurate solid phase and
its melt using probability density contour plots.  
Raising the temperature from 3 K to 4 K causes 
melting, so that each adsorption site will, after a sufficiently
long simulation run, have an equal probability of being occupied.
Further evidence for solidification comes from static structure factors.
Figure \ref{fig:ssffull} shows our
calculations at and immediately below the
commensurate solid density for the (01)
scattering direction.  The peaks at 
1.70 and 3.40 $\AA^{-1}$ are the
wave vectors expected for the first two Bragg scattering peaks for the 
$\sqrt 3 \times \sqrt 3$ solid.

\begin{figure}[htp]
\epsfxsize=\figwidth\centerline{\epsffile{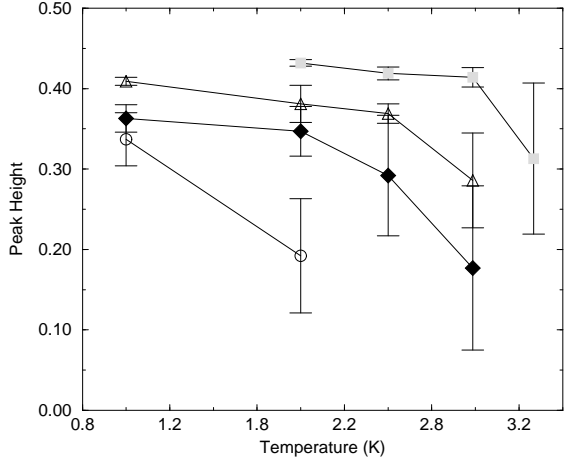}}
\caption{Temperature dependence of the peak height of the static
structure factor.  
The coverages are 0.0424 (circles), 0.0530 (filled diamonds), 
0.0566 (triangles), and 0.0636 atom/$\AA^2$ (filled squares).}
\label{fig:ssftemp}
\end{figure}

The temperature dependence of the static structure peaks
can be used to determine melting temperatures.  Figure \ref{fig:ssftemp}
shows $S(k=1.7 \AA^{-1})/N$ for several densities.  Melting is signaled by
a drop in the average peak height and large statistical 
fluctuations in peak values.  This is first observed at 2 K,
2.5 K, 3 K, and 3.33 K for 0.0424, 0.0530, 0.0566, and 0.0636 
atom/$\AA^{-2}$, respectively.  This density dependence of 
melting is in agreement with the experimental phase diagram, although 
our melting temperatures are somewhat higher than the experimental
values.  Heat 
capacity measurements
indicate the commensurate solid melts at 3 K, and the low density 
($\leq 0.045$ atom/$\AA^{-2}$) melting peaks are at about 1.5 K.  

\begin{figure}[htp]
\epsfxsize=\figwidth\centerline{\epsffile{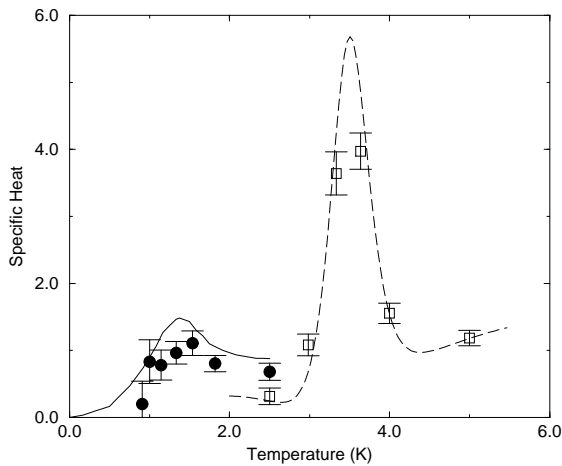}}
\caption{The heat capacity at 0.0353 (filled circles) and 
0.0636 atom/$\AA^2$ (squares).  The dashed line 
is a guide to the eye.  The solid line 
is the measured specific heat 
at 0.0367 taken from Ref. \protect \cite{greywall91}}
\label{fig:specheat}
\end{figure}

We have also calculated the temperature dependence of the
energy per particle for several densities.  These possess inflection
points that produce heat capacity peaks when the values are differentiated,
signaling melting.  Two sample
calculations of the specific heat are shown in Fig. \ref{fig:specheat}. 
At the commensurate density, we observe that melting occurs at about 3.5 K,
somewhat above the experimental value but consistent with the
static structure calculations.
As the density decreases, the melting temperature and peak height also
decrease.  At 0.0353 atom/$\AA^2$, the 
melting temperature is about 1.5 K, and the peak is much smaller and 
more rounded.

The binding energy $E_B$ of a single particle on the substrate may be 
easily calculated.  We find $E_B=-145.48 \pm 0.21$, which is comparable 
to, but slightly lower than, the estimated values of
$-141.75 \pm 1.50$ K from scattering\cite{thermo}
and $-142.33 \pm 1.97$ K from thermodynamic analysis\cite{elgin74}.  
Subtracting $E_B$ from the energy per particle at $\rho_c$,
we find the 
solid binding energy is $-2.21 \pm 0.20$ K.  This is lower than
either the energy per particle of the 2D liquid or solid determined in
Ref. \cite{bruch93}.

Having demonstrated that our simulation method can reproduce known features
of the monolayer, we now turn to the low density phase.  This
region is investigated by applying the
Maxwell common tangent construction to the low temperature values of the total 
energy to identify unstable regions at effectively
zero temperature.  The application of this
method to a system with a constant volume and 
varying particle number is described in Refs. \cite{pierce}.  
The free energy at nonzero temperatures 
is not directly accessible from PIMC calculations, but 
we may determine effectively zero-temperature energy values by a limiting
process \cite{pierce}.  The total free energy and total energy are the same at 
zero temperatures.

\begin{figure}[htp]
\epsfxsize=\figwidth\centerline{\epsffile{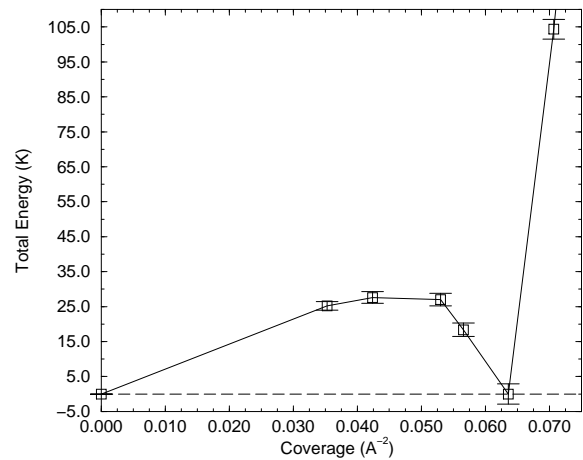}}
\caption{Total energy versus coverage.  For clarity, the energy values 
have been shifted by the line $Ne_0$, where $N$ is the number of particles, and
$e_0=-147.78 \pm 0.08$, the minimum energy per particle.  $e_0$ occurs at
the commensurate density.  The dashed
line is the gas-solid coexistence line.
}
\label{fig:engsqr3}
\end{figure}

The coexistence region between two stable phases is characterized by
an unphysical upward curvature of the total (free) energy.  In the 
thermodynamic limit, the energy values lie on a coexistence line.
Such an unstable region may be identified between zero coverage and
$\rho_c$ in the total energy
values shown in  Fig. \ref{fig:engsqr3}.  We have verified that all
these energy values have approached the zero-temperature limit
within error bars.  All intermediate energy values are
above the coexistence line.  
In the unstable region, the system
can phase separate into a zero density vapor and a commensurates solid
cluster.  The energy increase results from the
finite cost of creating the phase boundary.

\begin{figure}[htp]
\epsfxsize=\figwidth\centerline{\epsffile{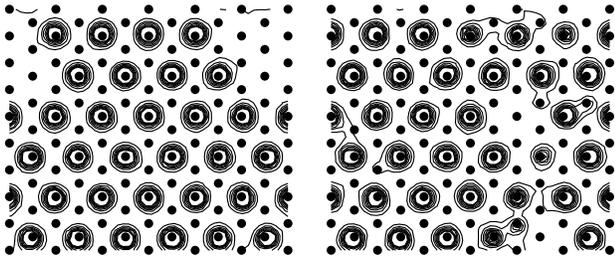}}
\caption{Probability distributions for 0.0566 atom/$\AA^2$ at
T=1.0 (left) and 2.5 K (right).
The filled circles give the locations of graphite potential minima.
}
\label{fig:distrib566}
\end{figure}

Contour plots of the probability density, shown in Fig. \ref{fig:distrib566}, 
provide direct evidence that both solid-vapor coexistence and
solid phases with vacancies occur.
At the lower temperature (left-hand side of Fig. \ref{fig:distrib566}),
the vacancies have coalesced.  Note that there is only one bubble in the
left-hand side of Fig. \ref{fig:distrib566} because of 
periodic boundary conditions.  The holes move very slowly at this 
temperature, producing
long equilibration times for condensation.  We thus calculated the
energy for this system with the vacancies initially separated
and then initially condensed.  The energy for condensed 
vacancies was lower.
At higher temperatures, the vacancies acquire enough kinetic energy
to leave the phase separated state and diffuse into the solid.  
As a result, vacancies can
become isolated.  This is illustrated at 2.5 K in the right-hand plot
of Fig. \ref{fig:distrib566}.  A series of probability distribution plots 
reveals that these vacancies move in the simulation, so the equilibration
problem encountered at 1.0 K is not present at this temperature.
We note that in our simulation 
there is still evidence of phase separation in contour plots
at 2.0 K for the density shown in Fig. \ref{fig:distrib566},
while experimental results seem to indicate
a transition at 1.5 K.  We have plotted probability contours for
densities as low as 0.0207 atom/$\AA^2$ and observe solid clusters at all 
densities.

We have also attempted to place a vacancy in a solid cluster surrounded
by vapor at 0.0424 atom/$\AA^2$ and 1.0 K.  
The vacancy was spontaneously expelled from the cluster during
thermalization, from which we conclude that
the solid clusters cannot also contain isolated vacancies when 
in equilibrium with the low density vapor.

Finally, we wish to discuss the arguments of Greywall and Busch (GB)
against solid clusters and in favor of the superfluid phase.
Their primary objection to solid-vapor
coexistence is that this should be signaled by
linear heat capacity isotherms for the entire region from
zero coverage up to the commensurate density.  Their
published data shows that for temperatures from 0.2 K to
0.5 K, the isotherms are linear only between 0.025 and 0.060 atom/$\AA^2$.  
At 0.1 K, the upper endpoint is about 0.055 atom/$\AA^2$.  
As a possible explanation, we
suggest that the departure from linearity below
0.025 atom/$\AA^2$ 
is caused by the presence of multiple finite-sized clusters.  At low densities,
solid clusters nucleate around surface defects.
Initially, there are many small metastable clusters with large 
perimeter-to-area
ratios.  Increasing the density increases the size of the clusters until
the surface is covered by a few large solid clusters with negligible
boundary effects.  Thus, the heat capacity exhibits linear 
behavior only after the solid clusters are sufficiently large
so that the perimeter-to-area ratio is small.  This
presumably occurs for coverages above 0.025 atom/$\AA^2$.  
GB have used a similar explanation
in their arguments for solid-liquid and liquid-gas coexistence in 
regions that do not have linear isotherms.

GB are lead to identify coverages near 0.04 atom/$\AA^2$ as
liquid based partly on simulation results for 2D helium on a flat
substrate, which were the most relevant calculations then
available.  As GB note, the large peak associated with 
the melting of the uniform commensurate solid phase first 
emerges at 0.04 atom/$\AA^2$.  2D helium is a liquid
near this density \cite{whitlock88}, suggesting that 
first layer coverages below 0.04 
may be liquid.  Unlike the purely 2D simulations, 
our calculations take the role of
substrate effects into account.  As we have shown, surface corrugations
push the density of the 
energy minimum up from about 0.04 on a flat substrate to 0.0636
atom/$\AA^2$ and produce solidification.  
GB also show that their heat capacity
results are in general agreement with a PIMC 
calculation for 2D superfluid helium \cite{ceperley89}, 
suggesting that there might be a superfluid
transition in the first layer.  We have
shown in Fig. \ref{fig:specheat} 
that the rounded heat capacities seen for low first layer densities
are produced by the melting of a solid cluster and are not associated with 
a superfluid transition.

This work was supported in part by the National Aeronautics and Space
Administration under grant number NAG3-1841.
Some of the calculations for this work were performed using the
facilities of the Supercomputer Computations Research Institute at 
Florida State University.

\end{document}